\documentclass[aps,prl,twocolumn,lengthcheck,groupedaddress,showpacs]{revtex4}
\usepackage{amsmath}
\usepackage{graphicx}
\usepackage{ifthen}
\usepackage{amsfonts}
\usepackage{amssymb}%
\def\n{\eta}

\def\be{\begin{equation}}
\def\ee{\end{equation}}
\newtheorem{theorem}{Theorem}

\newtheorem{Assumption}[theorem]{Assumption}

\newtheorem{Proposition}[theorem]{Proposition}

\begin{document}
\title{Aftershock identification}

\author{Ilya Zaliapin} 
\email{zal@unr.edu}
\affiliation{Department of Mathematics and Statistics,
University of Nevada, Reno, NV 89557-0084, Corresponding author.}


\author{Andrei Gabrielov} 
\email{agabriel@math.purdue.edu}
\affiliation{Departments of Mathematics and Earth and 
Atmospheric Sciences, Purdue University, West Lafayette, IN, 47907-1395}

\author{Vladimir Keilis-Borok}
\email{vkb@ess.ucla.edu}
\author{Henry Wong} 
\affiliation{Institute of Geophysics and Planetary Physics,
and Department of Earth and Space Sciences, University of California Los Angeles, 
3845 Slichter Hall, Los Angeles, CA 90095-1567} 
 
\date{December 7, 2007}

\begin{abstract}
Earthquake aftershock identification is closely related to the question 
``Are aftershocks different from the rest of earthquakes?''
We give a positive answer to this question and introduce a general statistical 
procedure for clustering analysis of seismicity that can be used, in particular, 
for aftershock detection.
The proposed approach expands the analysis of Baiesi and Paczuski 
[PRE, 69, 066106 (2004)] based on the space-time-magnitude nearest-neighbor 
distance $\n$ between earthquakes. 
We show that for a homogeneous Poisson marked point field with exponential marks, 
the distance $\n$ has Weibull distribution, which bridges our results with
classical correlation analysis for unmarked point fields.
We introduce a 2D distribution of spatial and temporal components of $\n$, 
which allows us to identify the clustered part of a point field.
The proposed technique is applied to several synthetic seismicity models and 
to the observed seismicity of Southern California. 
\end{abstract}

\maketitle
\newpage
\newpage

\section{Introduction}
Earthquake clustering is the most prominent feature of the observed seismicity.  
The centennial world-wide observations have revealed a wide variety of 
clustering phenomena that unfold in the time-space-magnitude domain (magnitude being 
the logarithmic measure of earthquake energy) and provide the most reliable and useful 
information about the essential properties of earthquake flow. 
Well-studied types of clustering include aftershocks, foreshocks, pairs of large earthquakes, 
swarms, bursts of aftershocks, rise of seismic activity prior to a large regional earthquake, 
switching of the global seismic activity between different parts of the Earth, {\it etc.}
Single clustering phenomena and their combination are an essential element 
of understanding the seismic stress redistribution 
and lithosphere dynamics \cite{stress}, as well as constructing empirical 
earthquake prediction methods and evaluating regional seismic hazard \cite{prediction}.

Baiesi and Paczuski \cite{BP} have developed an elegant framework for 
studying earthquake clustering by defining the pairwise earthquake distance $\eta_{ij}$ 
via the expected number of events in a particular time-space-magnitude domain 
bounded by events $i$ and $j$.   
These authors used the distance $\eta_{ij}$ to develop a tree-based statistical technique
for earthquake cluster analysis and established several scaling laws for the observed
earthquake clusters.

We expand here the approach of Baiesi and Paczuski \cite{BP} to demonstrate the existence of 
two statistically distinct subpopulations in the observed seismicity of Southern California:
One corresponds to a uniform, absolutely random flow of events while another to earthquake 
clustering.
The earthquakes from the clustering part, by and large, obey the conventional 
definitions of aftershocks \cite{AFT}.
Our analysis, therefore, provides an objective statistical foundation for 
aftershock identification that requires no prior clustering parameters like 
space-time windows traditionally used for aftershock detection \cite{AFT}. 

Our finding is supported by theoretical and numerical analyses of several seismicity
models, including ETAS \cite{ETAS}.
The main theoretical result is that for a homogeneous spatio-temporal Poisson field 
with independent exponential magnitudes, the distance $\eta$ has Weibull distribution,
the same distribution as the Euclidean nearest-neighbor distance for a homogeneous 
point field.
The proposed cluster detection technique is build upon the deviations of the observed 
nearest-neighbor distance $\eta$ from this theoretical prediction.  
The key element of the applied analysis is the 2D distribution of spatial and temporal 
components of $\eta$; this distribution clearly separates the clustered and non-clustered
parts of a point field.
\section{Distance between earthquakes}
Consider an earthquake catalog
$\{t_i,\,\theta_i,\,\phi_i,m_i\}_{i=1,\dots,N}$.
Each record $i$ describes
an individual earthquake with occurrence time $t_i$, 
position given by latitude $\theta_i$ and longitude 
$\phi_i$, and magnitude $m_i$; here, we do not consider
the depth.

For any two earthquakes $i$ and $j$ we define the time-space-magnitude
distance by
\begin{equation}
\label{nij}
n_{ij} = \left\{\begin{array}{rc}
C\,\tau_{ij}\,r_{ij}^d\,10^{-b\,(m_i-m_0)}, & \tau_{ij}\ge 0,\\
\infty,& \tau_{ij}<0.\end{array}\right.
\end{equation}
Here 
$\tau_{ij}=t_j-t_i$ is the earthquake intercurrence time; 
$r_{ij}$ surface distance;
$d$ is the fractal dimension of earthquake epicenters; and 
$b$ is the parameter of Gutenberg-Richter
relation (exponential fit to the magnitude
distribution): 
\be
\mathsf{P}\{m>x\} = 10^{-b\,(x-m_0)}I_{\{x>m_0\}}.
\label{GR}
\ee
%
Connecting each event with its nearest neighbor with respect
to the distance $n$ one obtains a time-oriented tree $\mathcal{T}$ whose 
root is the first event in the catalog.
Such trees were introduced and studied by Baiesi and Paczuski \cite{BP}.

It is readily checked that the space-time volume of a ball
of radius $C$ in metric $n$, $B_C:=\{(t,x,y,m)\,:\,n(t,r,m)<C\}$, is infinite due 
to heavy tails of the distance $n$ in time when $d>2$, in space when $d<2$,
and in both time and space for $d=2$.  
Hence, any such ball almost surely contains an infinite number 
of events from $N$ that prevents meaningful nearest-neighbor 
analysis.
To avoid this, we introduce the truncated distance   
\begin{equation}
\label{eta}
\eta_{ij}=\left\{\begin{array}{ll}
n_{ij},& t_{ij}\le t_0,\,r_{ij}\le r_0,\\
\infty,& {\rm otherwise.}\end{array}
\right.
\end{equation}
Choosing $t_0$ and $r_0$ large enough will ensure that the measures $\eta$ and $n$ 
are equivalent within a bounded spatio-temporal area. 
The {\it nearest-neighbor} distance is defined as
$\n^*_j:=\min_i\,\n_{ij}$. 
We will drop the subindices $ij$ or $j$ unless it is 
important which pair of earthquakes is considered.

\section{Main result: Poisson field}
\label{Poisson}
Consider a spatio-temporal marked point field $N$ 
with temporal component $t\in\mathbb{R}$, spatial component
${\bf x}\in\mathbb{R}^2$ and scalar marks $m$ that represent the earthquake 
magnitude.
Below we formulate our main result, starting with essential
assumptions about the field $N$.

\begin{Assumption}
\label{assumption}
{\rm
(i) $N$ is a homogeneous Poisson marked point field
with intensity $\lambda$.
(ii) Magnitude marks $m_i$ are independent of the field 
$(t_j,{\bf x}_j)$ and each other and have exponential
distribution \eqref{GR} with parameters $\tilde b$, $\tilde m_0$.
(iii) Let $f=b/\tilde b$ and 
$\mu_0=10^{\tilde b(\tilde m_0-m_0)}$ 
where $b$ and $m_0$ are the prior parameters of the Gutenberg-Richter
law (\ref{GR}) used in (\ref{nij}).}
\end{Assumption}

\begin{Proposition}
Under the Assumption \ref{assumption}, the nearest-neighbor distance 
$\eta^*_j$ has the following distribution, for large $\tau_0, r_0$:
\begin{equation}
\label{n_p}
\mathsf{P}\{\eta_j^*<x\}
=1-\exp\left(-\lambda\gamma\,\Psi\left(\frac{x}{\tau_0\,r_0^d\,\mu_0^f}\right)\right).
\end{equation}
Here $\gamma$ is independent of $x$ and we have
\begin{equation}
\label{fin_lim}
\Psi(w)\sim
\left\{\begin{array}{ll}
w,            & d<2,  f<1, \\
w\log w,      & d=2,  f<1, \\
w^{2/d},      & d>2,  d>2f,\\
w^{2/d}\log w,& d>2,  d=2f,\\
w^{1/f},      & d<2f, f>1, \\
w\log w,      & d<2,  f=1, \\
w(\log w)^2,  & d=2,  f=1,
\end{array}\right.
\end{equation}
where $\Psi(w)\sim \psi(w)$ stays for 
$\displaystyle\lim_{w\to\infty} \frac{\Psi(w)}{\psi(w)} = 1$.
\label{main}
\end{Proposition}
{\bf Proof} will be published elsewhere.

\begin{figure}[t]
\centering\includegraphics[width=.4\textwidth]{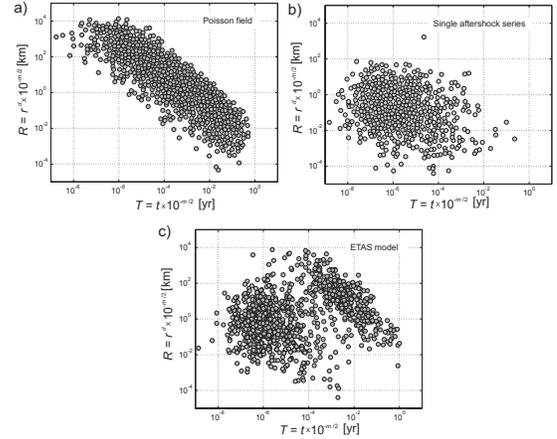}
\caption{Distribution of time and space components, $(T,R)$, of
the nearest-neighbor distance $\n$ for
homogeneous Poisson field with exponential magnitudes (a), 
single aftershock series obeying Omori law (b),
ETAS model (c).} 
\label{tx_model} 
\end{figure}

Proposition~\ref{main} implies that, for $b\ne\tilde b$, $d\ne2$,
and $d\ne 2f$, $\n^*$ has Weibull distribution. 
Furthermore, the distribution of $\n^*$ is independent of the magnitude threshold
$m_0$, when the latter is known (which is obviously the case in practice).
This facilitates analysis of data from different periods and 
regions that might have different $m_0$.

Let earthquake $i$ be the nearest neighbor for earthquake $j$,
that is $\n^*_j=\n_{ij}$.
We define, for arbitrary $0\le q \le 1$,
\begin{equation}
\label{pq}
T_{ij}:=\tau_{ij}\,10^{-b\,m_i\,q},\quad
R_{ij}:=r_{ij}^d\,10^{-b\,m_i\,(1-q)}.
\end{equation}
Obviously $\n^*_j=T\,R$ (without loss of generality, we assumed here 
$C=1$ and $m_0=0$) and Proposition \ref{main} implies 
that the distribution of the pair $(T,R)$ is concentrated along 
the line 
$\log_{10}T + \log_{10}R = x_{\rm m}$, 
where $x_{\rm m}$ is the mode of the distribution
\eqref{n_p},
while the level lines are of the form 
$\log_{10}T + \log_{10}R = $const.
Figure~\ref{tx_model}a illustrates this by showing the empirical 
distribution of the pairs $(T,X)$ for a Poisson homogeneous field 
with exponential magnitudes.

\section{Modeled seismicity}
\label{models}
Here we analyze numerically the distribution of nearest-neighbor
distances $\n^*_j$ for three point field models: 
(i) homogeneous Poisson marked field,
(ii) single self-excited aftershock series governed by Omori law, and 
(iii) ETAS model that combines the first two. 

The Epidemic Type Aftershock Sequence (ETAS) model was introduced 
by Y. Ogata \cite{ETAS}; it specifies a marked point process $N$ by
its conditional intensity at instant $t$ and spatial location
$(x,y)$: 
\begin{equation}
\label{ETAS}
\Lambda(t,x,y) = \Lambda_0 
+ \sum_{i\,:\,t_i<t} 10^{\,b\,m_i}\Lambda_T(\tau)\,
\Lambda_R(r), 
\end{equation}
where $\Lambda_0>0,\tau=t-t_i,r^2=(x-x_i)^2+(y-y_i)^2$, and
the temporal ($\Lambda_T$) and spatial
($\Lambda_R$) kernels are given by \cite{ETAS}
$\Lambda_T(t) = (t+c)^{-1-\epsilon_T}$,
$\Lambda_R(r) = (r+d)^{-1-\epsilon_R}$
with positive $c,d,\epsilon_T$ and $\epsilon_R$.
Magnitudes are drawn independently from the 
exponential distribution. 

A single aftershock series is a particular case of ETAS
model with $\Lambda_0$ replaced by $\delta(0,0,0)$ that
represents the mainshock; its magnitude is a model parameter.

\begin{figure}[h]
\centering\includegraphics[width=.4\textwidth]{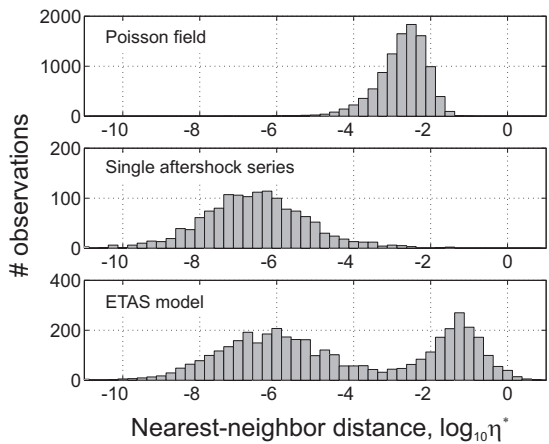}
\caption{Distribution of the nearest-neighbor distance $\n$
for homogeneous Poisson field with exponential magnitudes (top), 
single aftershock series obeying Omori law (middle),
ETAS model (bottom).} 
\label{eta_hist} 
\end{figure}

Figures \ref{tx_model} and \ref{eta_hist}  
show the distributions of $\eta^*$ and 
corresponding pairs $(T,R)$.
The Poisson model behaves as suggested by the Proposition~
\ref{main}.
For a single aftershock series, one observes almost symmetric 
scatter, which suggests that $T$ and $R$ are independent.
This is the most important difference from the Poisson model.
The ETAS distribution has two prominent ``modes'':
A scatter along $TR=$const. in the upper right part 
of the plot and an apparently independent scatter
closer to the origin.
Evidently, combining the homogeneous Poisson flow
and aftershock clustering we have combined as well
the corresponding modes of the $(T,\,R)$ distributions.

\section{Observed seismicity: Southern California}
We use a Southern California earthquake catalog produced by the Advance 
National Seismic System (ANSS) \cite{ANSS}, and consider earthquakes with 
magnitude $m\ge 2.0$ that fall within the square region bounded by
$122^{\circ}W$, $114^{\circ}W$, 
$32^{\circ}N$, $37^{\circ}N$
during January 1, 1984 - December 31, 2004.

\begin{figure}[h]
\centering\includegraphics[width=.4\textwidth]{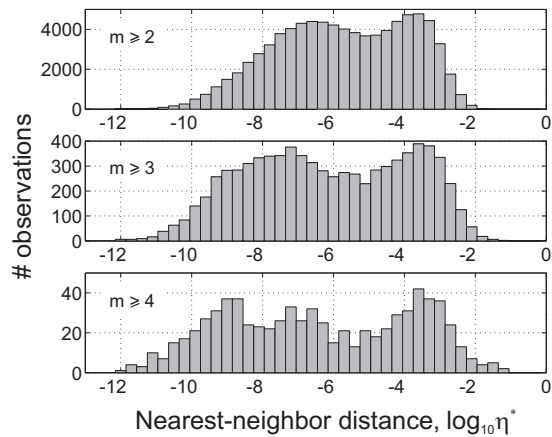}
\caption{Distribution of the nearest-neighbor distance $\n$ for the 
observed seismicity of Southern California during 1984-2004;
different panels correspond to different lower magnitude cutoffs.
Notice the bimodal structure with the same boundary between modes
at $\n\approx 10^{-5}$.} 
\label{CA_hist} 
\end{figure}

\begin{figure}[h]
\centering\includegraphics[width=.4\textwidth]{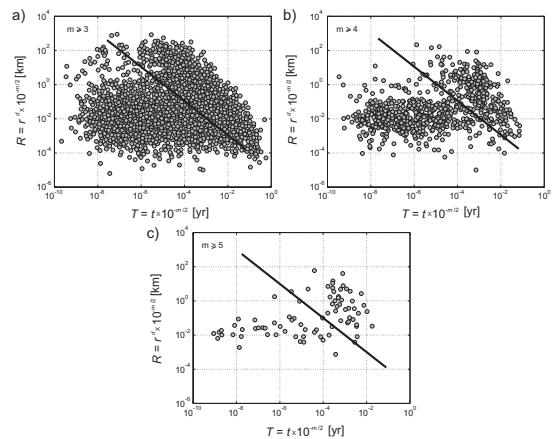}
\caption{Distribution of time and space components, $(T,R)$, of
the nearest-neighbor distance $\n$ for the observed seismicity
of Southern California during 1984-2004. 
Notice the bimodal structure; the location of a solid line 
$\log_{10}T+\log_{10}R=10^{-5}$ is the
same in all panels.} 
\label{tx_CA} 
\end{figure}

The empirical distributions of the nearest-neighbor distance $\n^*$
and its components $(T,\,R)$ are shown in Figs.~\ref{CA_hist},\ref{tx_CA}.
Both distributions are prominently bimodal reminiscent of that observed 
for ETAS model; they reveal existence of two statistically distinct 
earthquake populations.
One of them corresponds to $\log_{10}T+\log_{10}R \approx 10^{-3}$;
according to the Proposition \ref{main} it describes homogeneous 
(Poisson) seismicity.
The other population corresponds to $\log_{10}R\approx 10^{-2}$;
it corresponds to the aftershock clustering.

\begin{figure}[h]
\centering\includegraphics[width=.35\textwidth]{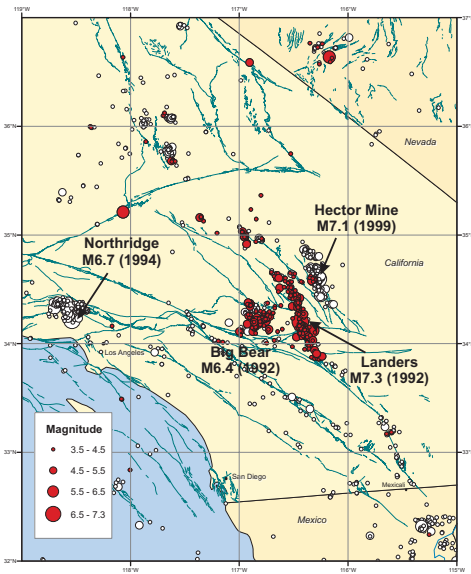}
\caption{(Color online) Aftershock identification for Landers earthquake 
(June 28, 1992, M7.3).
The figure shows all earthquakes that occurred after the Landers.
Shaded circles mark earthquakes identified as Landers' aftershocks;
open circles mark the rest of earthquakes. } 
\label{CA_aft} 
\end{figure}

To detect individual aftershocks, we fix a threshold $\eta_0$ and remove
all the links with $\eta_j^*>\eta_0$ from the tree $\mathcal{T}$.
This will result in the forest (set of trees)
$\mathcal{F}(\eta_0)=\{\mathcal{T}_i\}_{i=1}^{N(\eta_0)}$.
Each tree $\mathcal{T}_i$ in the forest corresponds to a single 
earthquake cluster: 
The distance between {\it linked} elements within any tree is smaller than 
that between any two elements from distinct trees.
Those clusters can be further analyzed in order to solve a particular
applied problem.
For example, aftershocks are often assumed to have smaller magnitude than 
the corresponding mainshocks \cite{AFT}.
Possible earthquake clusters observed prior to the mainshock are then 
called {\it foreshocks}.  
In this situation, it is natural to define $i$-th {\it mainshock} as the 
largest earthquake within the tree $\mathcal{T}_i$, and {\it aftershocks}
({\it foreshocks}) as the events from $\mathcal{T}_i$ that occurred later 
than (prior to) the mainshock. 
The results of this aftershock-detection procedure in California are shown in
Fig.~\ref{CA_aft}; here we used $\n_0=10^{-5}$ suggested by the distribution 
of $\n^*$ and $(T,X)$ (Figs.~\ref{CA_hist},\ref{tx_CA}).
The figure focuses on Landers earthquake, the largest one in California during 
the considered period.
The three groups of earthquakes are identified as aftershocks:
a) the prominent earthquake cluster in the immediate vicinity of the Landers' 
epicenter;
b) the ``secondary'' aftershocks after the Big Bear earthquake, M=6.4, which itself 
is the largest aftershock of Landers;
c) several earthquakes that occurred immediately after Landers
but at large distance from the latter. 
Such ``distant'' aftershocks present a special 
interest in many seismic studies. 
Both Northridge and Hector Mine aftershock clusters have not been associated
with Landers. 
We emphasize though existence of a distant Landers' aftershock close
to the future epicenter of Hector Mine. 

\section{Conclusion and Discussion}
We demonstrated the existence of statistically distinct clustered and 
non-clustered parts in the observed seismicity. 
This finding has important implications for various problems, aftershock detection 
being the most prominent one.
The physical interpretation of the reported separation as well as
its further applications will be considered in 
a forthcoming paper.

The current definition of the distance $\n$ remains {\it ad hoc}; a partial
justification for this choice is provided by our result on the distribution
for $\n^*$ (Proposition 2), which coincides with the Euclidean nearest-neighbor 
distance distribution for a homogeneous (unmarked) point field.
An analog of Proposition 2 is readily proven for any nearest-neighbor distance 
that depends multiplicatively on spatio-temporal point location and
multidimensional mark ${\bf m}$: $\eta=\tau\,r^d\,f({\bf m})$.
It would be interesting to see how alternative definitions of $\n$
will alter the applied part of the proposed clustering analysis.  

\acknowledgments
This study was partly supported by NSF, Grant ATM 0327558 and 
the Southern California Earthquake Center. 
SCEC is funded by NSF Cooperative Agreement EAR-0106924 and USGS Cooperative 
Agreement 02HQAG0008. 
The SCEC contribution number for this paper is 1137.

\end{document}